\documentclass[12pt,preprint]{aastex}
\newcommand\mone{$^{-1}$}
\def\gae{\mathrel{>\kern-1.0em\lower0.9ex\hbox{$\sim$}}}

\def\chandra{{\it Chandra }}

\def\xmm{{\it XMM-Newton }}

\def\ciaov{\hbox{\rm CIAO\thinspace v4.0 beta 2 }}
\def\caldb{\hbox{\rm CALDB\thinspace v3.4.1 }}
\def\sas{\hbox{\rm SAS\thinspace v7.1.0 }}

\def\kev{\hbox{$\rm\thinspace keV$}}

\def\yr{\hbox{$\rm\thinspace yr$}}

\begin{document}
\title{An Infrared Survey of Brightest Cluster Galaxies. II: 
Why are Some Brightest Cluster Galaxies Forming Stars?}
\author{Christopher P. O'Dea, Stefi A. Baum, George Privon, Jacob Noel-Storr}
\affil{Department of Physics, Rochester Institute of Technology,
  84 Lomb Memorial Drive, Rochester, NY 14623-5603}
\email{odea@cis.rit.edu}
\email{baum@cis.rit.edu}
\email{gcp1035@cis.rit.edu}
\email{jake@cis.rit.edu}
\author{Alice C. Quillen, Nicholas Zufelt, Jaehong Park}
\affil{Department of Physics and Astronomy,
  University of Rochester, Rochester, NY 14627}
\email{zufelt72@potsdam.edu}
\email{jaehong@pas.rochester.edu}
\email{aquillen@pas.rochester.edu}
\author{Alastair Edge}
\affil{Institute for Computational Cosmology, Department of Physics,
  Durham University, Durham DH1 3LE}
\email{alastair.edge@durham.ac.uk}
\author{Helen Russell, Andrew C. Fabian}
\affil{Institute of Astronomy, Madingley Rd., Cambridge,
  CB3 0HA, UK}
\email{hrr27@ast.cam.ac.uk}
\email{acf@ast.cam.ac.uk}
\author{Megan Donahue}
\affil{Michigan State University, Physics and Astronomy Dept.,
  East Lansing, MI 48824-2320}
\email{donahue@pa.msu.edu}
\author{Craig L. Sarazin}
\affil{University of Virginia, Department of Astronomy,
P.O. Box 400325, Charlottesville, VA 22904-4325}
\email{cls7i@mail.astro.virginia.edu}
\author{Brian McNamara}
\affil{University of Waterloo, Department of Physics and Astronomy,
200 University Avenue West, Waterloo, Ontario, Canada N2L 3G1}
\email{mcnamara@uwaterloo.ca}
\author{Joel N. Bregman}
\affil{University of Michigan, Physics  Dept.,
Ann Arbor, MI 48109}
\email{jbregman@umich.edu}
\author{Eiichi Egami}
\affil{Steward Observatory, University of Arizona, 933 North Cherry Avenue, Tucson, AZ 85721 }
\email{eegami@as.arizona.edu}

\begin{abstract}
Quillen et al.(2007) presented an imaging survey with the  {\it Spitzer Space Telescope}
of 62 brightest cluster galaxies with optical line emission located in the cores of
X-ray luminous clusters.
They found that at least  half of these sources have signs of excess infrared emission.
Here we discuss the nature of the IR emission and its implications for cool core clusters.
The strength of the mid-IR excess emission correlates with the luminosity of the optical
emission lines.	 Excluding the four systems dominated by an AGN, the excess mid-infrared
emission in the remaining brightest cluster galaxies is likely related to star formation.
The mass of molecular gas (estimated from CO observations) is correlated with the IR
luminosity as found for normal star forming galaxies. The gas depletion time scale is
about 1 Gyr.
The physical extent of the infrared excess is consistent with that of  the optical emission
line nebulae. This supports the hypothesis that the star formation occurs in molecular
gas associated with the emission line nebulae and with evidence that the emission line
nebulae are mainly powered by ongoing star formation. We find a correlation between
mass deposition rates (${\dot M}_X$) estimated from the X-ray emission and the star
formation rate estimated from the infrared luminosity.	The star formation rates are
1/10 to 1/100 of the mass deposition rates suggesting that the re-heating of the ICM
is generally very effective in reducing the amount of mass cooling from the hot phase
but not eliminating it completely.
\end{abstract}


\section{Introduction}

The hot T$\sim 10^{7-8}$~K X-ray emitting gas is currently thought to
constitute the bulk of the baryonic mass in rich clusters of galaxies.
An important aspect of the overall physics of the intracluster medium (ICM)
concerns the central
regions of clusters ($r \lesssim 10 - 100$~kpc), where the inferred ICM
densities and pressures in some cases are sufficiently high that cooling to
$T \lesssim 10^4$~K can occur on time scales shorter than the cluster lifetime
	(e.g., \citealt{cowie77, fabian77, edge92}).
These ``cooling core'' clusters often exhibit intense optical emission-line
nebulae associated with the centrally dominant (cD) galaxies at their centers,
together with blue continuum excess emission, and the strength of these effects
appears to correlate with the cooling rate or central pressure of the X-ray
emitting gas \citep{heckman81, johnstone87, romanishin87, mcnamara92, mcnamara93,
crawford92, crawford93, allen95}.

The previous paradigm pictured the ICM as a relatively simple place where gas
cooled and slumped in towards the center of the cluster in a cooling flow
with mass accretion rates of hundreds of solar masses per year (e.g., \citealt{fabian94b}).
However, X-ray spectroscopy with XMM-Newton and Chandra has failed to find evidence
for gas at temperatures below about one-third of the cluster virial
temperature (e.g., \citealt{kaastra01, tamura01, peterson01, peterson03, peterson06}).
The limits on the luminosity of the intermediate temperature
gas imply reductions in the inferred mass accretion rates by factors of 5-10.
Recent theoretical models indicate that intracluster conduction,
combined with an episodic heat source in the cluster core, such as an AGN
or star formation, are candidates for explaining both the X-ray emission
from cluster cores and the optical emission-line phenomena associated
with the cores with these rapid-cooling spectra (e.g., \citealt{ruszkowski02, 
voigt02, fabian02, narayan01}). 
One widely considered possibility is that an important source of
heat in the ICM are bubbles driven by radio galaxies (e.g., \citealt{baum91,
tucker97, soker02, bohringer02, kaiser03, omma04, dunn05, dunn06, birzan04, rafferty06}) 
which halts the cooling
of the gas.  The ICM now appears to be a very dynamic place where heating and cooling
processes vie for dominance and an uneasy balance is maintained.
Since these same processes may operate during the process of
galaxy formation, the centers of clusters of galaxies provide low redshift
laboratories for studying the critical processes involved in galaxy formation and
supermassive black hole growth. At the present time, the main questions are 
(1) How much gas is cooling out of the ICM? (2) How much star formation is ongoing? 
(3) What is the impact of the gas and star formation on the central BCG?

As little mass is needed to power the AGNs at the center
of bright cluster galaxies the only way to remove cooled
gas from the ICM is to form stars.  Measurements of the star formation
rate in cluster galaxies can therefore provide constraints
on the efficiency of cooling, the fraction of gas
that cools and so the needed energy input to prevent the remainder of the
gas from cooling.  It is also possible that the ICM in cluster galaxies is not
in a steady state or experiences periods of enhanced cooling and star formation
and periods of relative activity when cooling is prevented.
Star formation and associated
supernovae also provide a source of mechanical energy, though
this is not sufficient to match the X-ray radiative energy
losses \citep{mcnamara06}.

ISO observations detected the cluster S\`ersic 159-03 \citep{hansen00}.
Recent Spitzer observations have demonstrated that star formation is common
in cooling core BCGs \citep{egami06a,donahue07b,quillen07}. An infrared excess is found
in about half of the sample of 62 BCGs studied by \citet{quillen07} (Paper I). 
In this paper we discuss the results of \citet{quillen07}. We examine correlations
in the data and discuss the implications for star formation in BCGs and
the balance of heating and cooling in the ICM. 
Specifically we 
search for correlations between star formation rates,
radio, H$\alpha$, CO and X-ray luminosities and mass deposition rates estimated
from the X-ray observations.
In this paper all luminosities
have been corrected or computed to be consistent with a Hubble constant
$H_0 = 70$ Mpc$^{-1}$ km~s$^{-1}$ and a concordance cosmology
($\Omega_M=0.3$ and flat).

\section{Comparison data}

The properties of the BCG sample are discussed by
\citet{quillen07}. 
Comparison data for the BCGs in our sample are listed
in Table 1 by \citet{quillen07}. 
When available, this Table lists 
X-ray (primarily ROSAT 0.1-2.4 keV), radio (1.4 GHz), 
and H$\alpha$ luminosities (from long-slit spectra and SDSS data)
and $[$OIII$]$(5007$\rm \AA$)/H$_\beta$ flux ratios.
Brightest cluster galaxies can host both star formation and an active
galactic nucleus.  X-ray luminosities
provide a constraint on the mass in and radiative losses
from the hot ICM.  The H$\alpha$ recombination
line is excited by emission from hot stars produced during
formation or from an AGN.  
We note emission lines are detected in $\sim 10-20\%$  of typical optically
selected BCGs, $\sim 30-40\%$ of X-ray selected BCGs, and almost  
$100\%$ for BCGs in cooling core clusters 
\citep{donahue92,crawford99,best07,edwards07}.
To discriminate between the presence of an
AGN and star formation we have sought a measure of the
hardness of the radiation
field through the $[$OIII$]$(5007$\rm \AA$)/H$\beta$ optical line ratio.  Fluxes in
the radio also provide a constraint on the properties of the AGN.
Below we discuss 
star formation rates estimated using infrared luminosities derived 
from aperture photometry also listed in Paper I, molecular gas
masses estimated from CO observations and mass deposition rates
measured from X-ray observations.
The statistical tests for correlations between
the various quantities are given in Table \ref{tab:corr}.

\section{Estimated star formation rates}

\begin{figure*} 
\plotone{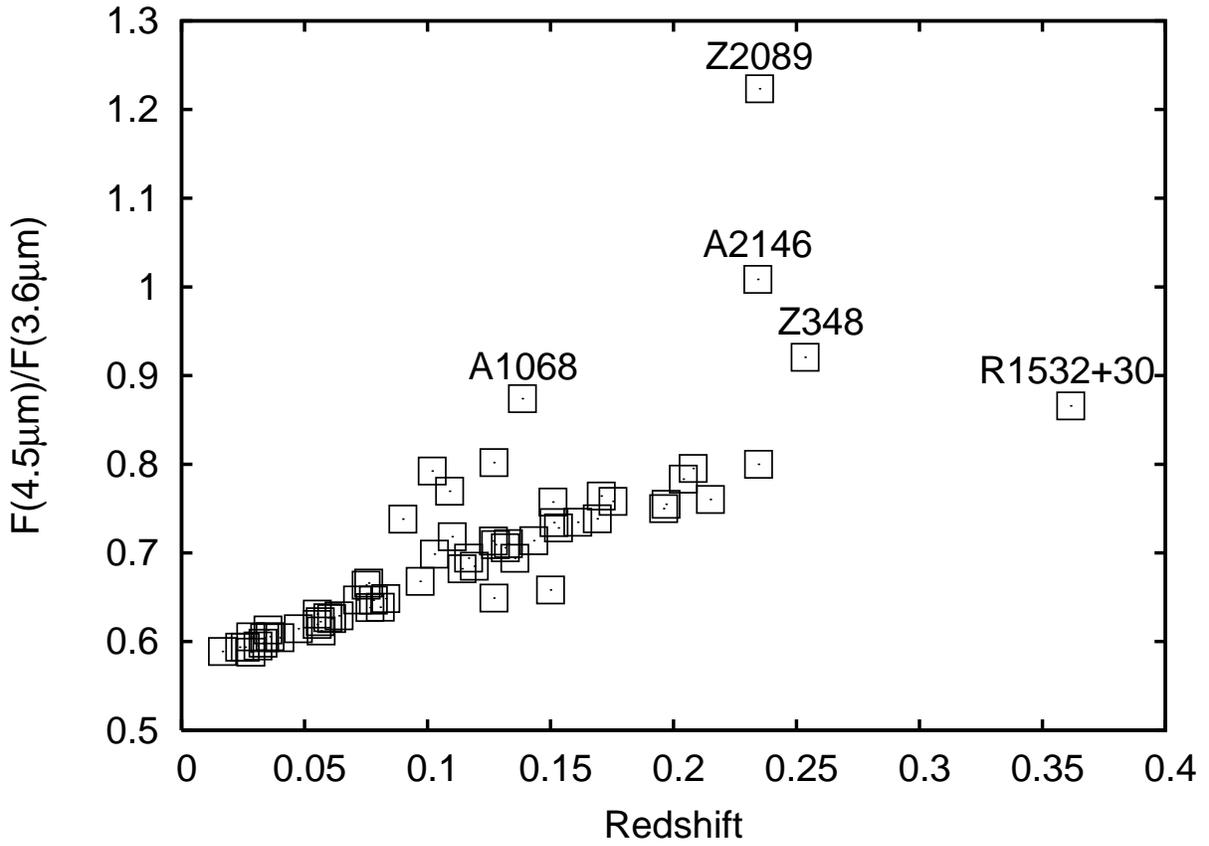}
\caption{Flux ratio $F_{4.5}/F_{3.6}$ vs. redshift. The labeled objects show 
evidence for
the presence of an optically luminous type 2 AGN. 
\label{fig:43redshift}}
\end{figure*}

If the infrared luminosity is powered by star formation, we can use the IR luminosity
to estimate a star formation rate (SFR) (e.g., \citealt{bell03,calzetti08}). But first we need
to consider whether some sources have a contribution to the IR from a Type II AGN
with an optically bright accretion disk. 
\citet{quillen07} identified Z2089,A1068, A2146  as likely to have an AGN
contribution based on red 4.5/3.6 micron color, unresolved nucleus
seen in IRAC color maps and high [OIII]/H$\beta$ flux ratio.
R0821+07 was flagged as possibly similar as it has  an unresolved
nucleus in IRAC color map and a high [OIII]/H$\beta$. It also has
a remarkably red 8.0/5.6 micron color similar to a Seyfert 2 with an embedded dusty AGN. 
In Figure~\ref{fig:43redshift} we plot the ratio of  4.5$\mu$ and 3.6$\mu$m fluxes 
against redshift (data from Paper I).
The clear trend seen is as expected for a passive stellar population
but with a few notable exceptions. The sources with strong [OIII]
(Z2089, A1068, and A2146) lie above the trend as do A2055
and A2627, that show evidence for a BLLac continuum in optical spectra
\citep{crawford99}. The two galaxies that lie below the trend are Z2072 and Z9077 and
are among the fainter objects in our sample. Z9077 was the only
object not detected at 24 microns.
The four that are between z=0.09 and  and z=0.15 and lie slightly above the trend are 
A1885, A2055, A2627 and R0352+19.  It's not obvious why these 4 sources lie above 
the trend or if this is significant R0352+19 and R0821+07 are quite red in the 
8/5.8 micron color,   and  R1532+30 and Z348 are pretty red in 8/5.8 but do not stand out in the 
4.5/3.6 color vs. z plot.  Thus, the combination of diagnostics (4.5/3.6 micron color,
red unresolved nuclear source, and high [OIII]/H$\beta$ ratio) identifies some sources
with a strong AGN contribution. The remaining objects are likely to be free of strong 
AGN contamination. 

Previous optical and  UV observations have found evidence for significant
star formation in the BCGs in cool core clusters \citep{johnstone87,romanishin87, mcnamara89, 
mcnamara93, mcnamara04, mcnamara04b, hu92, crawford93, hansen95, allen95, smith97, 
cardiel98, hutchings00, oegerle01, mittaz01, odea04,hicks05,rafferty06}.
Table 1 of \citet{quillen07} lists [OIII]/H$\beta$ ratios for most of the 
BCGs. Except for the few which may host an AGN, the ratios are consistent with 
the gas being ionized by hot stars. 

In Table \ref{tab:selected} we present the estimated infrared
luminosities from \citet{quillen07} and the estimated star formation
rates.
The SFR rate can be estimated from the infrared luminosity with
equation 5 by \citet{bell03};
\begin{equation}
\psi(M_\odot \ {\rm yr}^{-1})
=      A \left({ L_{\rm IR}\over L_\odot}\right) ( 1 + \sqrt{10^9 L_\odot/L_{\rm IR}}).
\label{eqn:SFR}
\end{equation}
Here the constant $A = 1.57\times 10^{-10}$
for $L_{\rm IR} > 10^{11}L_\odot$ 
and $A=1.17\times 10^{-10}$ at lower luminosities. 
The SFRs  are in the range of about 1 to a few tens of
$M_\odot$~yr$^{-1}$.  
The objects with SFR above about 50 $M_\odot$~yr$^{-1}$ are
likely AGN dominated. 

In Table~\ref{tab:SFR_comp} we list available SFRs in different
wavebands. We see that there is dispersion in the estimated SFRs.
However because of the effects of dust and geometry we do not necessarily expect
agreement between SFRs estimated in the IR vs. the UV/optical. 
Much of the variation can be accounted for by aperture
mismatch, differences in assumptions about star formation history
i.e., burst vs. constant star formation, extinction, and perhaps differences in the
amount of dust available to re-radiate in the FIR.  Note that Abell 1068 
and A2146 show large discrepancies between our FIR SFR and the U-band SFR and both
are flagged as possible AGN. Given the expected dispersion, the rough agreement 
between the star formation rates is consistent with the IR emission being dominated
by star formation.  


\subsection{Caveat - the Dust-to-Gas Ratio}

The derived SFR might be underestimated if the cold gas  in
the BCGs has a low dust-to-gas ratio. This might be the case if the gas has cooled from
the hot ICM and if the dust was destroyed while in the hot phase and there has 
not been sufficient time to form dust at the levels typically seen in normal star forming galaxies. 
However, there are several arguments against a low gas-to-dust ratio. (1) The observations of 
H$_2$ (e.g., \citealt{donahue00, edge02, hatch05, jaffe05, egami06b, johnstone07}) 
and CO \citep{edge01, salome03, salome04} associated with the BCG optical emission line
nebulae require the presence of significant amounts of dust to shield
the molecular gas. (2)  Dust is clearly seen in the optical emission line nebulae in cool
core clusters (e.g., \citealt{sparks89, sparks93, mcnamara92, donahue93, koekemoer99}).
(3) Studies of the nebulae in the BCGs of cool core clusters
suggest the presence of  dust-to-gas ratios consistent with Galactic values
\citep{sparks89, sparks93, donahue93}. (4)  Theoretical arguments suggest that
dust could form quickly inside cool clouds \citep{fabian94a, voit95}.

\section{ Comparison between infrared luminosity and X-ray luminosity }

\begin{figure*} 
\plotone{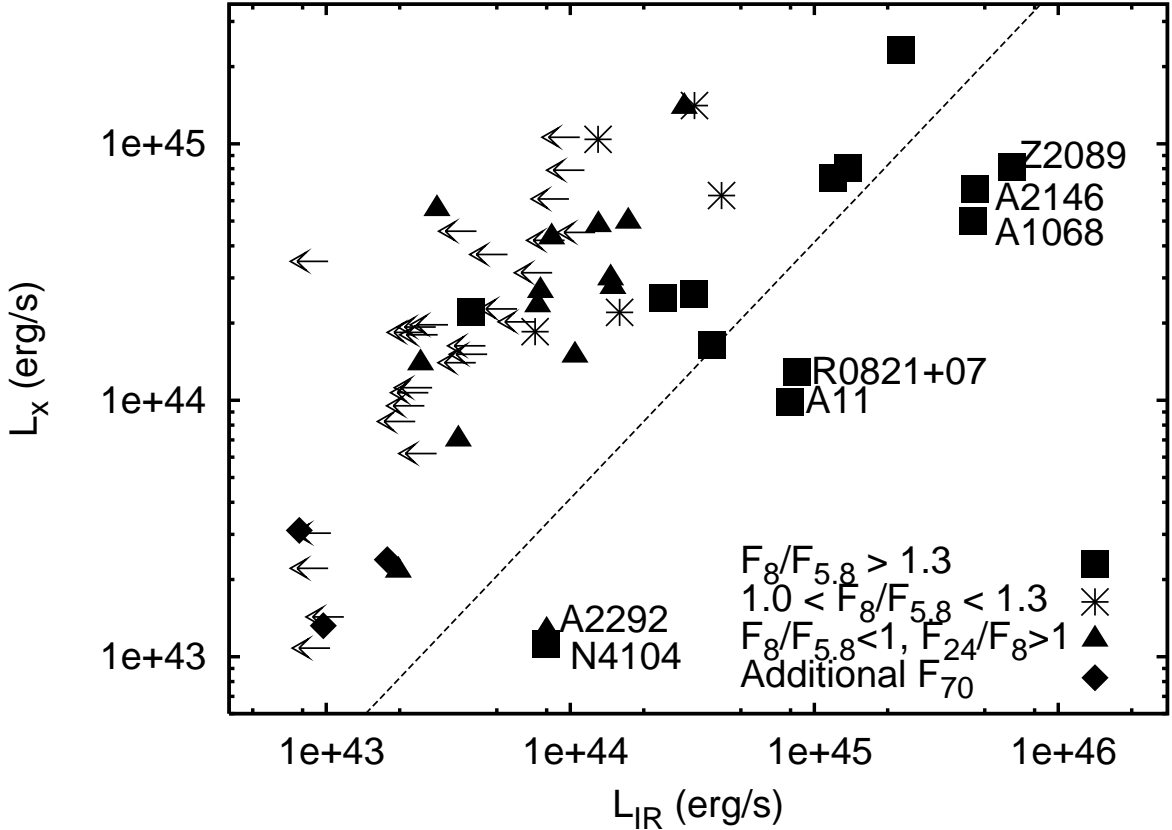}
\caption{X-ray luminosity vs. estimated infrared luminosity.
X-ray and infrared luminosities are listed in Tables 1 and 3 of Paper I.  
Solid squares are our reddest group with 8 to 5.8$\mu$m flux ratio  
greater than $1.3$.
The intermediate group with flux ratios between 1.0 and 1.3 are plotted
as stars.   
Solid triangles have 8 to 5.8 $\mu$m flux ratio less than 1 but
24 to 8 $\mu$m flux ratio above 1.  Solid diamonds are galaxies
with both 24 to 8 and 8 to 5.8 $\mu$m flux ratios less than 1
but have been detected at 70$\mu$m.
Flux ratios are computed using photometry listed in Table 2 
in Paper I. 
Upper limits on the infrared luminosity are shown by arrows.
We find a weak correlation between X-ray and infrared luminosity.
The dashed line  is the kinetic energy injection rate predicted
from a star forming population due to supernovae as a function of the infrared
luminosity.   We confirm that kinetic energy from supernovae cannot
account for the X-ray radiative energy losses in most cooling
flows.
\label{fig:lxvslir}}
\end{figure*}

\begin{figure*} 
\plotone{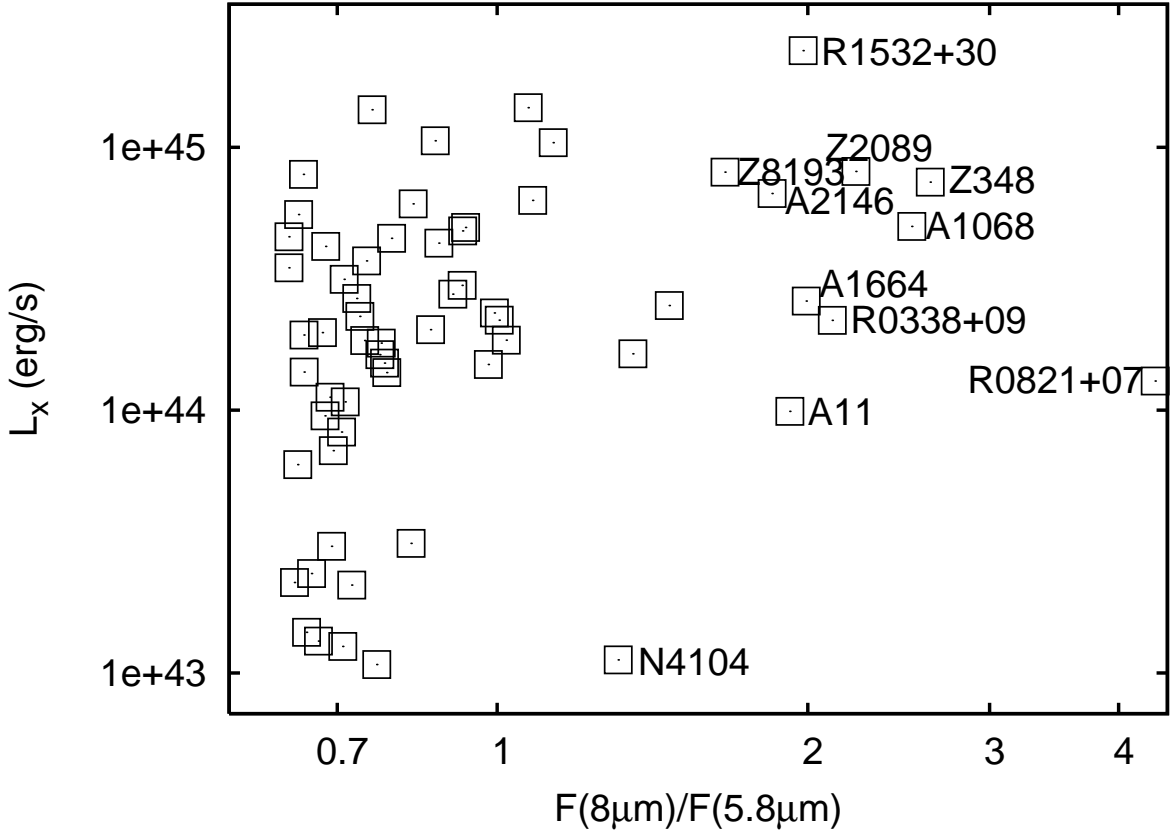}
\caption{
X-ray luminosity vs. 8 to 5.8$\mu$m flux ratio 
(data taken from Tables 1 and 2 in Paper I).   
Most red objects with $F_{8\mu}/F_{5.8 \mu {\rm m}} > 1$ 
have X-ray luminosity $L_X > 10^{44} $ erg/s.
\label{fig:lxvs85}}
\end{figure*}

The integrated X-ray luminosity of a cluster is dependent on the combination
of its core and larger scale structure. As such any correlation between
this global property and the properties of the BCG may indicate
an underlying link, particularly as our sample from its selection will
favor cool cores. Therefore we plot the X-ray luminosity of the host cluster 
(listed in Table 1 of Paper I) 
against estimated infrared luminosities for all BCGs with color
ratio $F_{8 \mu{\rm m}}/ F_{5.8\mu {\rm m}} > 0.75$ in Figure \ref{fig:lxvslir}.
In Figure \ref{fig:lxvs85}
we show X-ray luminosities compared to the color
$F_{8 \mu {\rm m}}/F_{5.8 \mu {\rm m}}$.
This study covers a much larger range in X-ray luminosity
than \citet{egami06a}. 
We see that BCGs with higher IR luminosity and redder 8 to 5.8$\mu$m colors 
(indicating an IR excess)  tend to have higher X-ray luminosities.
Though, there are many objects with high X-ray luminosity ($L_X > 10^{44}$ erg/s), 
which do not have an IR excess. 

It is interesting to compare the kinetic energy injected by supernovae
(from a star formation rate consistent with the infrared luminosity) to
the energy radiated in X-rays.  
\citet{leitherer99} estimate
a mechanical energy of about $10^{42}$erg~s$^{-1}$
normalized for a star formation rate of $1M_\odot$ yr$^{-1}$.
These conversion factors have been used to estimate the mechanical
energy due to supernovae as a function of infrared luminosity.
This relation is shown as a dashed line in Figure \ref{fig:lxvslir}.  
We see that there are a few BCGs for which there may be sufficient mechanical energy
to resupply the X-ray luminosity. 
However, in general, for the sample as a whole, we confirm the finding of 
previous studies (e.g., \citealt{mcnamara06}) that mechanical energy input 
from supernovae is not sufficient (by a few orders of magnitude)
to account for the current radiative energy losses of the intracluster
medium as a whole or in the core.

\section{Comparison to radio luminosity}

\begin{figure*} 
\plotone{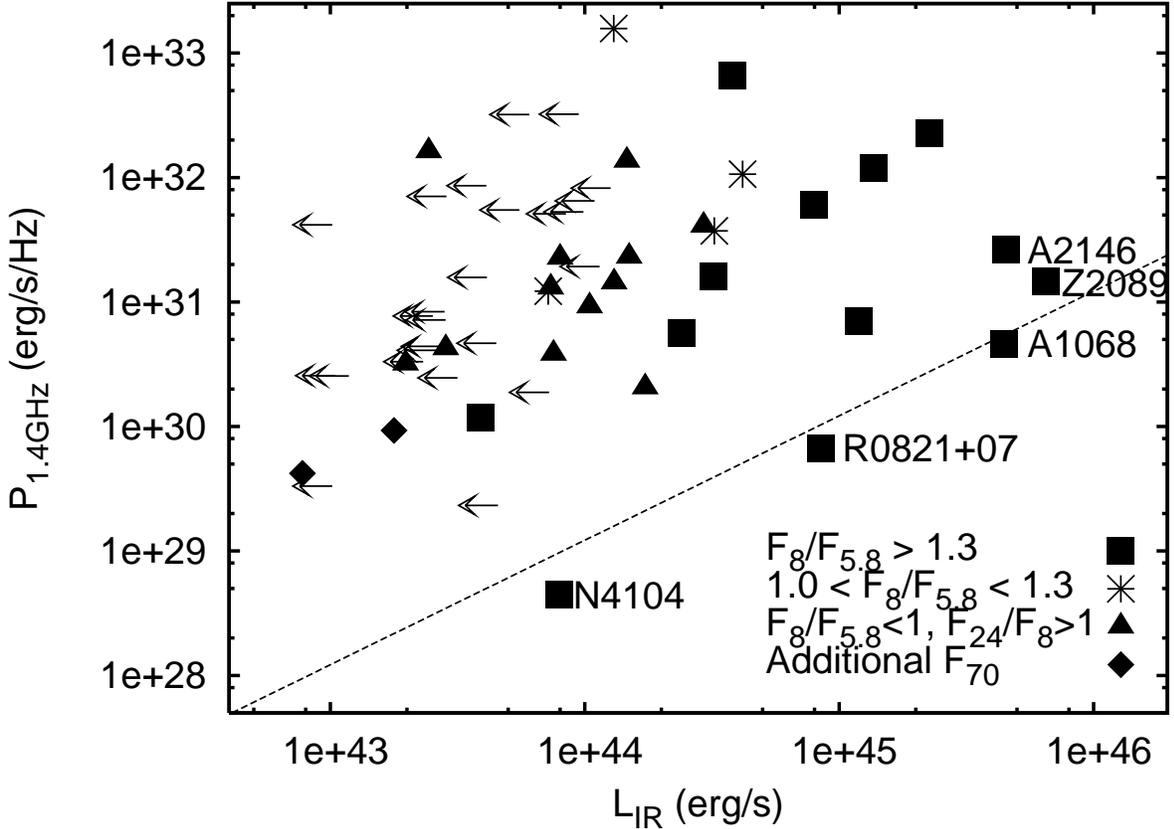}
\caption{Radio luminosity at 1.4 GHz (listed in Table 1 of Paper I)
vs. estimated infrared luminosity (Table~\ref{tab:selected}.
The radio-infrared correlation (equation \ref{eqn:rir})
for star forming objects is shown as a line on the lower right.  The radio
fluxes are much higher than expected from the radio-IR correlation appropriate
for star forming late type galaxies.  This is not unexpected since many of these
objects contain radio cores and in some cases even double lobed jets.
We find a weak correlation between the radio luminosity at 1.4 GHz and the
infrared luminosity. The point types are as given in Figure \ref{fig:lxvslir}.
\label{fig:lradiolir}}
\end{figure*}

We find a modest (almost 3$\sigma$)  correlation between the infrared luminosity
and the radio luminosity at 1.4 GHz (as we show in Figure \ref{fig:lradiolir}).
We compare the radio fluxes to those appropriate for star forming objects with
a solid line on the lower right in Figure \ref{fig:lradiolir}.
The radio-IR relation for star forming objects
(equation 3; \citealt{bell03})
\begin{equation}
\left({ L_{1.4 \rm GHz}\over {\rm erg~cm^{-2}~s^{-1} Hz^{-1}}}\right)
  =
\left({L_{\rm IR} \over   3.75\times10^{12+q}
\ {\rm erg~cm^{-2}~s^{-1}~Hz^{-1}}}\right)
\label{eqn:rir}
\end{equation}
where $q$ is a logarithmic index.
We have used the mean value $q=2.34$ by \citet{yun01}.

On Figure \ref{fig:lradiolir},
the majority of radio fluxes are well above this relation.
The three objects below the line are (from left to right)
NGC4104, R0821$+$07, A1068.  NGC4104 is nearer than the other objects 
in the survey and it is possible that the  H$\alpha$ flux and radio flux density have
been underestimated. 
The other two clusters (Abell 1068, R0821$+$07) have
$F_{8 \mu {\rm m}}/F_{5.8 \mu {\rm m}}>1.3$ and have
unresolved red sources seen in the IRAC color maps and so
are likely to be dominated by an AGN.

Thus, the BCGs (independent of whether they have an IR excess) tend to have  
radio emission which  is dominated 
by that produced by an AGN. Based on hot IR colors and high [OIII]/H$\beta$ ratios it appears
that only 4 of the BCGs host a  Type II AGN with a luminous accretion disk \citep{quillen07}.
Thus, either the  AGN in most of the BCGs are currently turned off, or 
they are accreting in a low luminosity mode. The (weak) correlation between
radio and IR luminosity may be a consequence of the correlation between
mass accretion rate and SFR \S~\ref{sec:mdot}; i.e., the cooling gas 
feeds the AGN and makes gas available for star formation. In addition,
the ratio of mechanical energy in the radio source outflow to  the radio luminosity
can vary by about 3 orders of magnitude \citet{birzan04}. Thus, the radio luminosity
can be a poor measure of the impact of the radio source on its environment. 

\section{Comparison to H$\alpha$ luminosity}

\begin{figure} 
\plotone{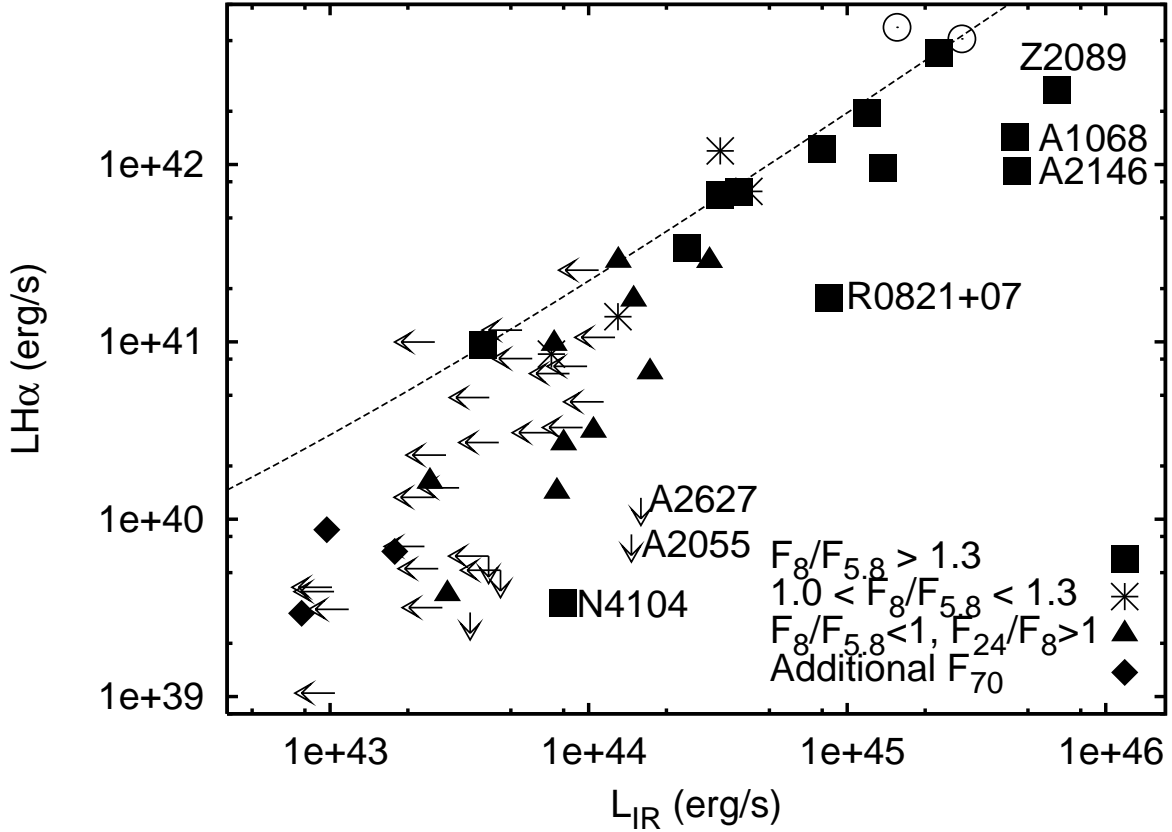}
\caption{Observed H$\alpha$ Luminosity (listed in Table 1 of Paper I)
vs Infrared Luminosity
estimated from the 8$\mu$m and 24.4$\mu$m fluxes.
The data for 2 BCGs from \citet{egami06a} are shown as open circles.
The Kennicutt relation inferred from observations of star forming galaxies
relating H$\alpha$ luminosity to star formation rate
is plotted as a line.  
We have divided the line by a factor of 2.8 
to back-out the reddening correction since our H$\alpha$ luminosities are
uncorrected for reddening. 
The H$\alpha$ fluxes are consistent with
the estimated infrared luminosities and star formation.  As is
true for Figure \ref{fig:lxvslir} the point types depend on
the 8 to 5.8$\mu$m color.
We suspect that some of the H$\alpha$ luminosities are lower than expected
because the apertures used to measure them were smaller than
those used to measure the infrared fluxes.
\label{fig:lalphalir}}
\end{figure}

\begin{figure} 
\plotone{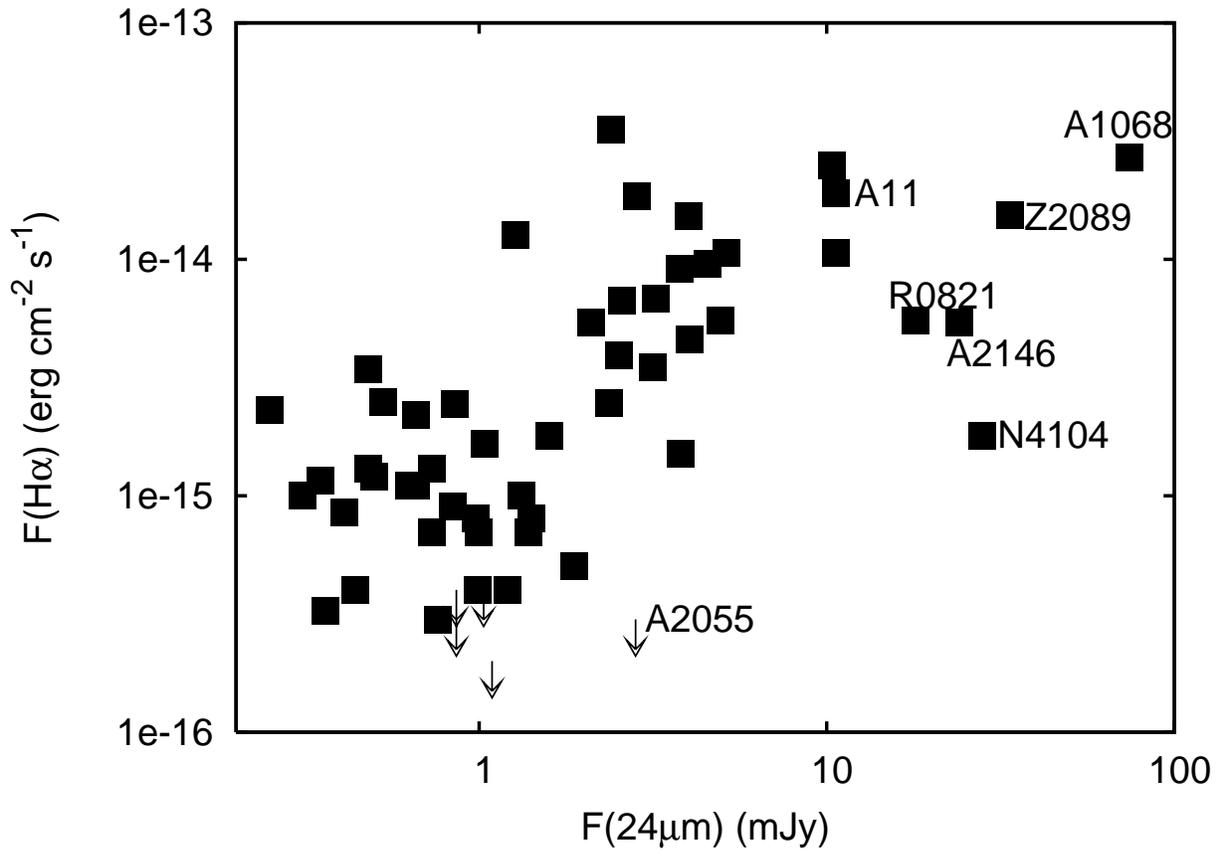}
\caption{Observed H$\alpha$ flux vs 24 $\mu$m flux listed
in Paper I. Arrows denote upper limits.
\label{fig:havs24}}
\end{figure}

We compare the H$\alpha$ luminosities from limited aperture spectroscopy 
to the infrared luminosities in Figure \ref{fig:lalphalir} finding a strong 
correlation between the two.
We also see a correlation in H$\alpha$ flux vs. 24 micron flux density (Figure
\ref{fig:havs24}). 
These correlations show that the H$\alpha$ and infrared emission arises from
the same or a related power source. We suggest that the dominant power source for
the H$\alpha$ and infrared emission is star formation.
This is consistent with previous evidence that the optical emission line  nebulae
are mostly powered by UV photons from young stars with a possible secondary
contribution from another mechanism  (e.g., \citealt{johnstone88, allen95, voit97, crawford99,
odea04, wilman06, hatch07}).
The H$\alpha$-SFR law relating the star formation rate
to the H$\alpha$ luminosity
\begin{equation}
SFR \ (M_\odot~{\rm yr}^{-1}) = { L({\rm H}\alpha) \over 1.26 \times 10^{41} {\rm erg~s^{-1}}}
\end{equation}
\citep{kennicutt98} is shown as a dashed line on the plot.
We have scaled the line down by a factor of
2.8 because our H$\alpha$ luminosities are uncorrected for reddening.
We see in Figure \ref{fig:lalphalir} that the points tend to lie
a factor of a few below this line, i.e., the
observed luminosity in H$\alpha$ is lower than that expected
from the estimated infrared luminosity. The discrepancy is larger at lower X-ray luminosity
$L_X < 10^{44}$ ergs/s. Our H$\alpha$ luminosities are taken mainly from
spectroscopy with a long slit of width $1\farcs 3$ \citep{crawford99} or the 3\arcsec\  diameter
fibers of the SDSS. Narrow band H$\alpha$+[NII] images and IFU observations
give angular sizes for 11 of our BCGs and calibrated H$\alpha$+[NII] fluxes for 6 sources
\citep{heckman81,heckman89,cowie83,baum88,mcnamara04b,wilman06,donahue07a,hatch07}.
We find that the nebulae are all larger than the spectroscopic apertures, with a median
size of $7\farcs 1$ (geometric mean of major and minor axes).
The total fluxes determined from the narrow band images and IFU spectroscopy 
are larger than
those from \citet{crawford99} or SDSS, with a median ratio of 1.4. Thus, it seems likely
that aperture effects contribute to the H$\alpha$ deficit, though larger samples
with narrow band imaging are needed to determine whether this can explain the 
whole effect.
Additionally, strong absorption of the	H$\alpha$ (relative to normal star forming
galaxies) could also contribute to the H$\alpha$ deficit.
However, the possibility remains that star formation is not the only power source
for the H$\alpha$ and IR emission. If this turns out to be the case, it would suggest 
that there is an additional  source of energy which heats the dust, but does not ionize the gas.
Such an energy source would help to explain the observed optical line ratios
\citep{voit97} and bright H$_2$ emission \citep{edge02, jaffe01,jaffe05}.

\section{Comparison to molecular gas mass }
\label{sec:co}

We have compiled  molecular mass data from
\citet{edge01},	 \citet{salome03},
and Edge, in preparation.
This subsample consists only
of objects that have been surveyed for	and detected in CO (1-0).
The inferred molecular gas masses range from $\sim 10^9$ to $\sim 10^{11}$ M$_\odot$.
Spitzer IRS spectra of the star forming BCG in Z3146 detect strong molecular 
hydrogen lines from warm H$_2$ with an estimated  mass of $\sim 10^{10}$ 
M$_\odot$ \citep{egami06b}.  This provides confirmation that the molecular gas 
masses can be very large in these BCGs.

\begin{figure*} 
\plotone{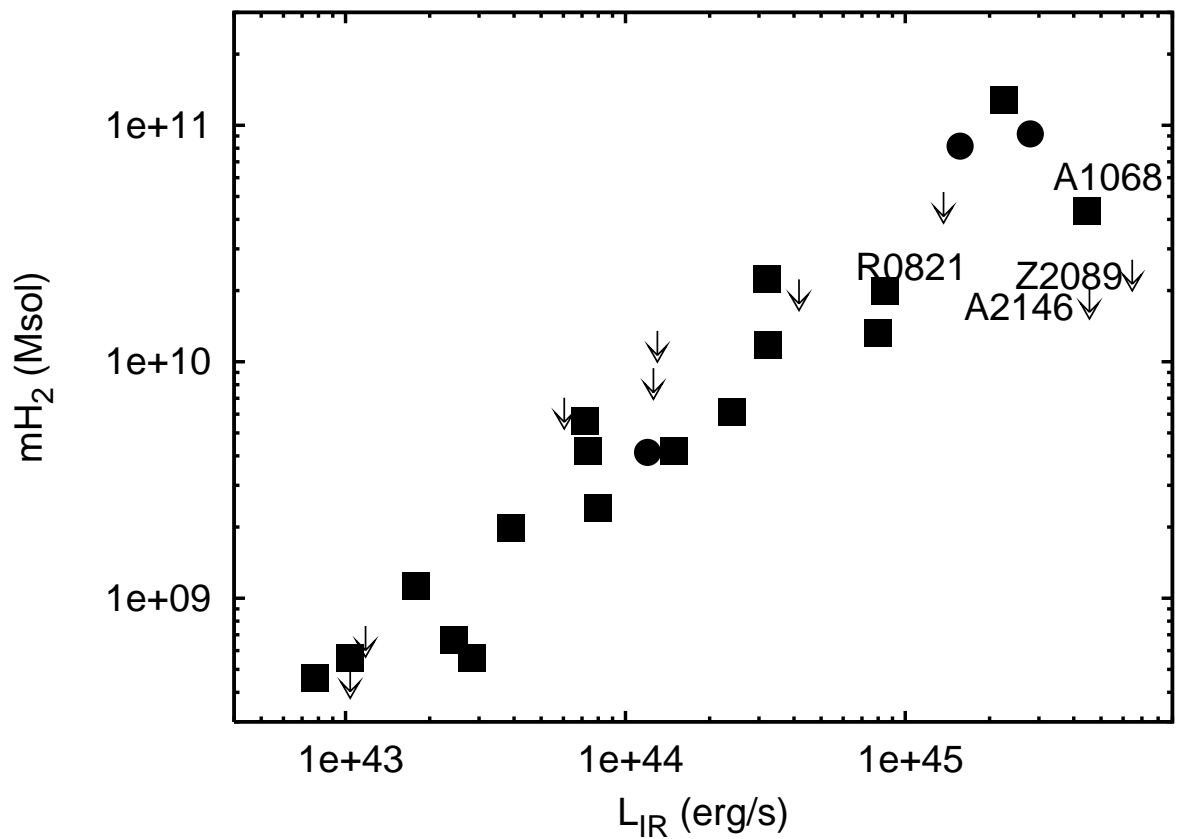}
\caption{Correlation of H$_2$ mass from CO luminosity and $L_{\rm IR}$. 
 A1835 and Z3146, discussed by \citet{egami06a}, and A2597 discussed
by \citet{donahue07b} are shown as filled circles. 
We find a strong correlation 
and as such consider the relation between CO and star formation.  
Upper limits are shown as arrows. 
The 4 objects thought to host AGNs are labeled.
Two of these have IR luminosity higher than expected from
their molecular gas mass.
\label{fig:coplots1}}
\end{figure*}

\begin{figure*} 
\plotone{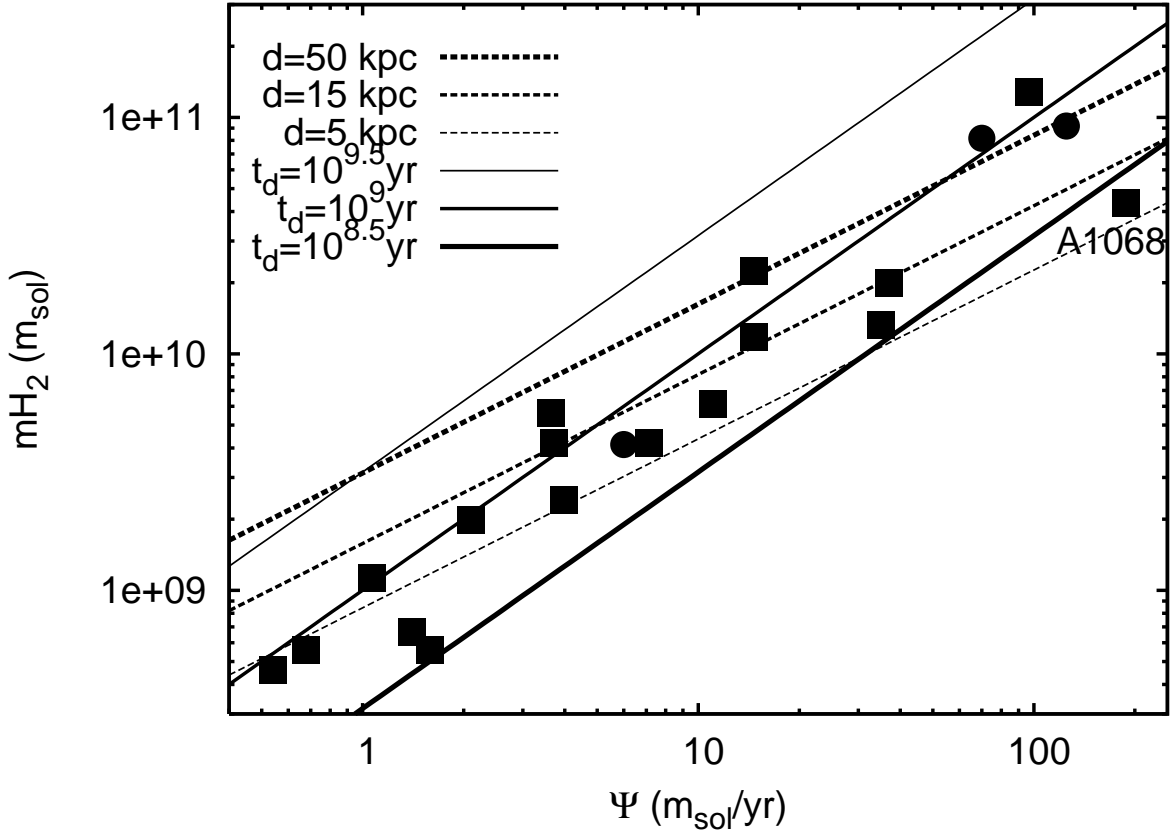}
\caption{Correlation of H$_2$ mass from CO luminosity and SFR.
2 BCGs, A1835 and Z3146, discussed by \citet{egami06a}, are shown as filled
circles. 
The dotted lines are taken from Eq. 4 by \citet{kennicutt98} using different 
values for the diameter of the star forming region. See the legend at the upper left.
The diameter of the star forming region
tends to be larger for more luminous objects.
Solid lines are computed assuming gas depletion time scales
of $10^{8.5}, 10^9$ and $10^{9.5}$ years. See the legend at the upper left.
\label{fig:coplots2}}
\end{figure*}

We note that a correlation between integrated molecular gas mass and H$\alpha$
luminosity in BCGs has been found by \citet{edge01} and \citet{salome03}.
In Figures \ref{fig:coplots1} and \ref{fig:coplots2}, we plot the molecular mass against
our estimated infrared luminosity and star formation rate
(listed in Table \ref{tab:selected}).
As found in normal star forming galaxies (e.g., \citealt{young86, kennicutt98}),
we see a correlation between measured molecular gas mass and both
the IR luminosity and the star formation rate in the BCGs.
The ratio of molecular gas mass to SFR gives a gas depletion time scale which is
roughly 1 Gyr. The gas depletion time scale is roughly constant over a range of
two orders of magnitude in molecular gas mass and SFR.
Our value of $\sim 1$ Gyr is in good agreement with the mean value of $\sim 2$ Gyr
found in normal star forming galaxies  by \citet{young86} which have molecular
gas masses in the range ($\sim 10^9$ to $10^{10}$ M$_\odot$).
The long life time of the molecular gas in these BCGs is in contrast to the much shorter
cooling times for the gas over a range of temperatures. The hotter phases cool on 
times of $\sim 10^6 - 10^8$ yrs \citep{peterson06}, while the molecular gas cools on even
shorter time scales \citep[e.g.,][]{jaffe01}. 
Given that clusters are relatively young (perhaps 4-6 Gyr since
last major merger) it is possible that there may have been insufficient time for a
complete  steady-state (cooling leads to cold gas leads to star formation) to be set up.

\subsection{The Size Scale of the Star Formation and its Relation to the
Optical Emission Line Nebulae}
\label{sec:co2}

In nearby galaxies there is an empirical relation
between star formation rate per unit area
and molecular gas surface density.  This relation can be
described in terms of a Schmidt-Kennicutt law
\citep{kennicutt98}
\begin{equation}
\left({\dot{\Sigma}_{\rm SFR}  \over
    M_\odot~{\rm yr}^{-1}~{\rm kpc}^{-2} }\right)
  =   2.5\times 10^{-4}
\left({\Sigma_{\rm gas} \over M_\odot~{\rm pc}^{-2}}\right)^{1.4}
, \label{eqn:schmidt}
\end{equation}
where $\Sigma_{\rm gas}$ is the surface density of molecular and atomic
gas and $\dot{\Sigma}_{\rm SFR}$ is the star formation rate per unit area.
We can use this relation to estimate the size scale of the star forming
region.
We make the assumption that the star formation is 
distributed in a region of area $d_{\rm kpc}^2$ where $d_{\rm kpc}$ is a diameter
in kpc and
the surface density $\Sigma_{\rm gas} ={M_{\rm H_2} / d_{\rm kpc}^2}$ and
where $M_{\rm H_2}$ is the molecular gas mass.
Applying this to equation \ref{eqn:schmidt},
we find a relation between molecular mass and star formation rate;
\begin{equation}
\left({M_{\rm H_2} \over M_\odot}\right)
=3.7 \times 10^8 ~ d_{\rm kpc}^{0.57}~
\left({SFR \over M_\odot \ {\rm yr}^{-1}}\right)^{0.71}.
\label{eqn:cosfr}
\end{equation}
We have shown this line
in Figure \ref{fig:coplots2} computed for diameters
$d_{\rm kpc}=5$, 15, and 50.
We see that the data are consistent with a Schmidt-Kennicutt law,
but the diameter of the star forming region is not well constrained. 
The diameter of the star forming region tends to be larger for more
luminous objects as expected if the diameter is proportional to 
$L_X$/${\dot M}_X$. 
Previous studies have shown that the Schmidt law
predicts the star formation rates within a factor of a few for galaxies
over a wide range of morphologies and star formation rates,  including
starbursts galaxies \citep{kennicutt98}.
The previous study of 2 BCGs by \citep{mcnamara06} suggested
that the star formation law holds even in BCGs. However, Figure~\ref{fig:coplots1}
shows that at high  molecular gas masses $M_{\rm H_2} > 10^{10}$ M$_\odot$, some
BCGs show inferred diameters $\gae 50$ kpc which are much larger than suggested
by the sizes of the emission line nebulae.

The sources with the largest estimated star formation regions are R1532+30,
A1664, and Z8197 with estimated star formation region size scales of 70, 50 and 30 kpc,
respectively, estimated using the Schmidt type star formation law.
None of these is well resolved, all have FWHM near the diffraction
limit of 7\arcsec\ at 24 microns.
For R1532+30 at z = 0.36, the FWHM corresponds to a size of 35 kpc.
This is below the estimated size of the star forming region, $R\sim 70$kpc.
Likewise for A1664, and Z8197, with redshifts of 0.128,	 and 0.114,
the FWHM corresponds to about 15 kpc and again this exceeds the estimated
size scale by a factor of 2-3.	The size scale estimates using the star
formation law are probably a factor of a 2-3 too large for these objects.
Those with the smallest estimated star formation regions are A85, A262, A2052,
and NGC4325  with estimated regions of	smaller than 5 kpc.
At a redshift of 0.0551,0.0166,	 0.0351, 0.0259, 7\arcsec\
(diffraction limit at 24 microns) corresponds to 7, 2.3, 5 and 3.6 kpcs, respectively.
The objects with the smallest estimated regions are the nearest
and so can be resolved in the IRAC images.
For A85, the star forming region could be the unresolved source at 8 microns
that is south east of the brightest cluster galaxy nucleus.
The brightest cluster galaxy is resolved at 8 microns.
For A262 and NGC4235, the brightest cluster galaxy is the source
of the 24 micron emission and is resolved both at 24 and 8 microns,
consistent with the estimate for the star forming region size of a few kpc.
For A2052 the brightest cluster galaxy also hosts star formation in
its nucleus. The emission is unresolved at 24 microns but resolved at 8 microns.
This is consistent with the estimated size of the star forming region
of a few kpc.
Except for the case of A85 the estimated sizes of the star forming
regions of a few kpc are consistent with the sizes estimated from
the images. In summary, the Kennicutt-Schmidt law gives sizes which are  generally 
consistent with those estimated from the images for the small and average sizes, 
though the largest sizes seem to be too large by factors of 2-3.

We note that the emission line nebula in cool cores tend to have a bright 
central region with a diameter of order 10 kpc (e.g., \citealt{heckman89}),
with fainter gas extending to  larger scales (e.g., \citealt{jaffe05}) which is comparable
to the inferred size of the star formation region. 
Observations of extended HI absorption in the emission line nebula of 
A2597 suggests that the optical nebulae are photon bounded and are the 
ionized skins of cold atomic and molecular clouds \citep{odea94}.
In addition, molecular hydrogen has been found associated with emission line
filaments in some BCGs 
(e.g., \citealt{donahue00,  edge02, hatch05, jaffe05, egami06b, johnstone07}).
Interferometric CO observations show molecular gas associated with 
the emission line nebula in A1795 (e.g., \citealt{salome04}).
Also, HST FUV images show FUV continuum from young stars
associated with the emission line nebulae in A1795 and A2597 \citep{odea04}. 
Thus, the spatial association of the FUV, the CO, the H$_2$, and the optical  emission
line nebulae suggest that the star formation occurs in molecular gas which lies
in the optical emission line nebulae.

Our estimated size scale of $d\sim 15$ kpc for the star formation
region could be biased.   Objects that have larger and more diffuse
star formation regions would have had larger molecular gas masses
and so would have been detected.  Similarly, BCGs with lower and more concentrated
star formation regions might have been missed.

\section{The Connection between Star Formation and the Properties of the Hot ICM}\label{sec:mdot}

We use archival \chandra and \xmm observations to calculate X-ray
inferred mass deposition rates and cooling times for 14 of the selected clusters. 
We required at least 15,000 counts from the source in each observation to generate
reliable deprojected spectra.  This restricted the cluster sample to
11 with suitable \chandra archive data and 3 with \xmm archive data 
(R0338+09, R2129+00 and Abell 115).

The \chandra data were analyzed using \ciaov with \caldb provided by
the \chandra X-ray Center (CXC).  The level 1 event files were
reprocessed to apply the latest gain and charge transfer inefficiency
correction and filtered for bad grades.	 Where available, the
improved background screening provided by VFAINT mode was applied.  The background light curves
of the resulting level 2 event files were then filtered for periods
affected by flares.  For the nearer clusters ($z<0.3$), background
spectra were extracted from blank-sky background
data sets available from the CXC and cleaned in the same way as the
source observations. The normalizations of these cleaned background files were 
scaled to the count rate of the
source observations in the 9--12\kev\ band.  For more distant clusters,
background spectra were extracted from suitable, source-free regions
of the source data sets.

The \xmm MOS data were reprocessed using the \texttt{emchain} task from
\xmm \sas to generate calibrated event files from the raw
data.  Cosmic ray filtering was applied by selecting only PATTERNs 0--12 and 
bright pixels and hot columns
were removed by setting FLAG$==$0.  Soft proton flares were removed by
generating a light curve for photons of energy $>10$\kev, where the
emission is dominated by the particle-induced background, and
rejecting high background periods.  Blank-sky background spectra were
produced using the blank-sky background event files available from
the \xmm Science Operations Center and
calibrated according to \citet{read03}.

Spectra were extracted in concentric annuli centered on the X-ray surface
brightness peak with a minimum of 3000 counts in each annulus.
\chandra spectra were analyzed in the energy range 0.5--7.0\kev and
\xmm spectra in the range 0.3--10 keV.	Suitable response files (ARFs
and RMFs) were calculated and grouped together with the source
spectrum, binned with a minimum of 30 counts.

Deprojected temperature and density profiles were calculated using a
Direct Spectral Deprojection method \citep{sanders07,sanders07b},
which creates 'deprojected spectra' using a model independent
approach, assuming only spherical geometry.  Instead of correcting for
projection by combining a series of models, this new method subtracts
the projected spectra from each successive annulus to produce a set of
deprojected spectra.

The resulting deprojected spectra were analyzed in XSPEC 11.3.2
\citep{arnaud96}. Gas temperatures and densities were found by fitting
each spectrum with an absorbed single-temperature plasma model
(\texttt{phabs(mekal)}).  The redshift was fixed to the value given in
Table 1 of Paper I and the absorbing column density was fixed to the
Galactic values given by \citet{kalberla05}. The temperature, abundance and model
normalization were allowed to vary.  We used the deprojected
temperature and density to determine the cooling time of the gas at
each radius.  The cooling radius was defined to be the radius within
which the gas has a cooling time less than $7.7\times10^9\yr$,
corresponding to the time since $z=1$.

\begin{figure*} 
\plotone{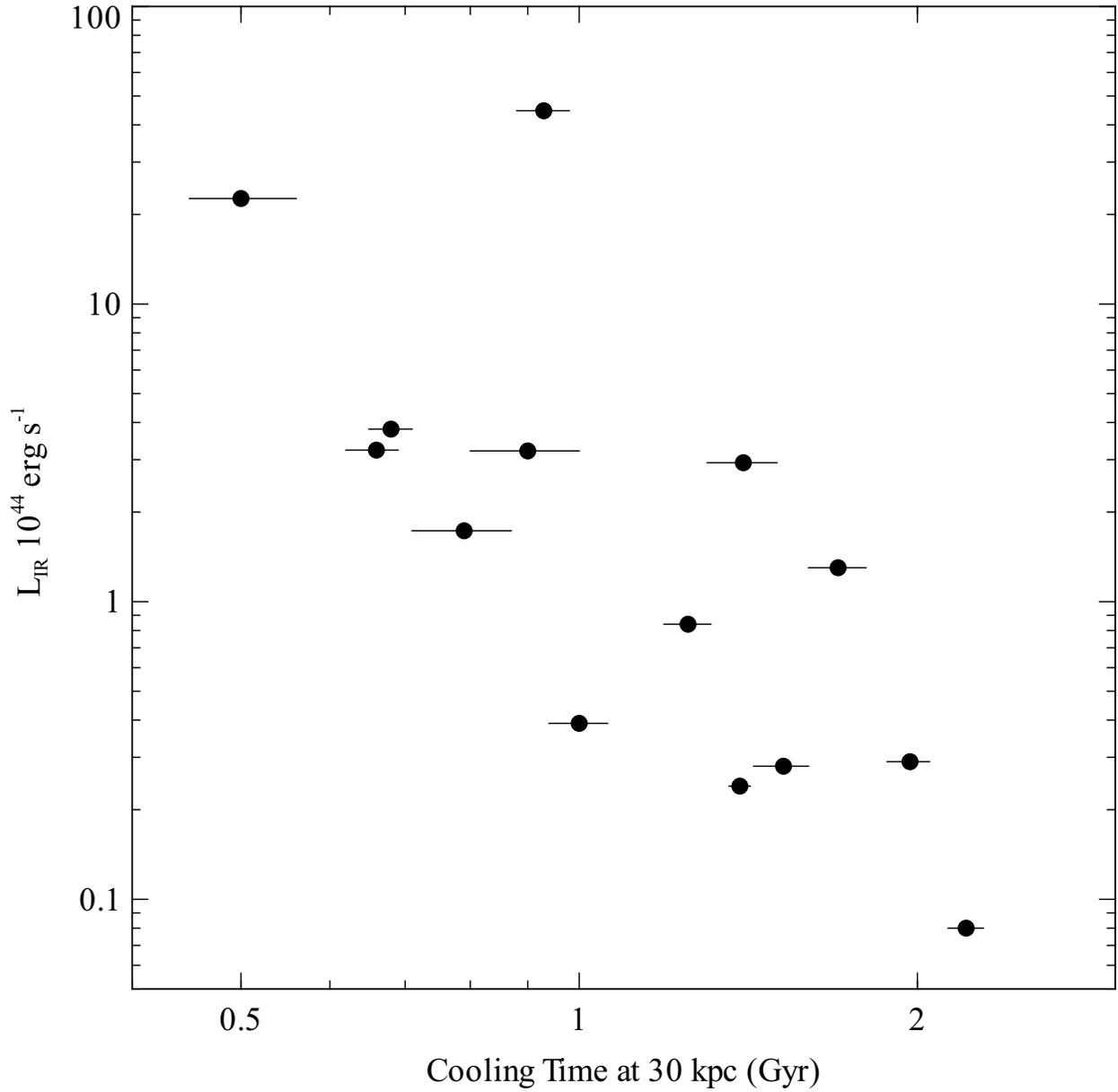}
\caption{IR luminosity vs. X-ray derived cooling times at a radius of 30 kpc. 
BCGs have higher IR luminosity in clusters with shorter cooling times. 
\label{fig:tcool}}
\end{figure*}

In Figure \ref{fig:tcool} we plot the Infrared Luminosity vs. the cooling time
at a radius of 30 kpc. We see that BCGs with shorter cooling times have higher
IR luminosities consistent with the results of \citet{egami06a}. This is
consistent with the hypothesis that the clusters with shorter cooling times
have higher star formation rates which result in higher IR luminosity.  

We calculate two different measures of the mass deposition rate.
$\dot{M}_I$ is a maximum mass deposition rate, calculated from

\begin{equation}
\dot{M}_I=\frac{2{\mu}m_H}{5k_B}\frac{L(<r_{cool})}{T(r_{cool})}
\end{equation}

\noindent where ${\mu}m_H$ is the mean gas mass per particle and the luminosity
was determined over the energy range 0.01--50 keV.  L($<r_{cool}$) is directly proportional to the
energy required to offset
the cooling and $\dot{M}_I$ is a measure of the mass deposition rate
if heating is absent.

\begin{figure*} 
\plotone{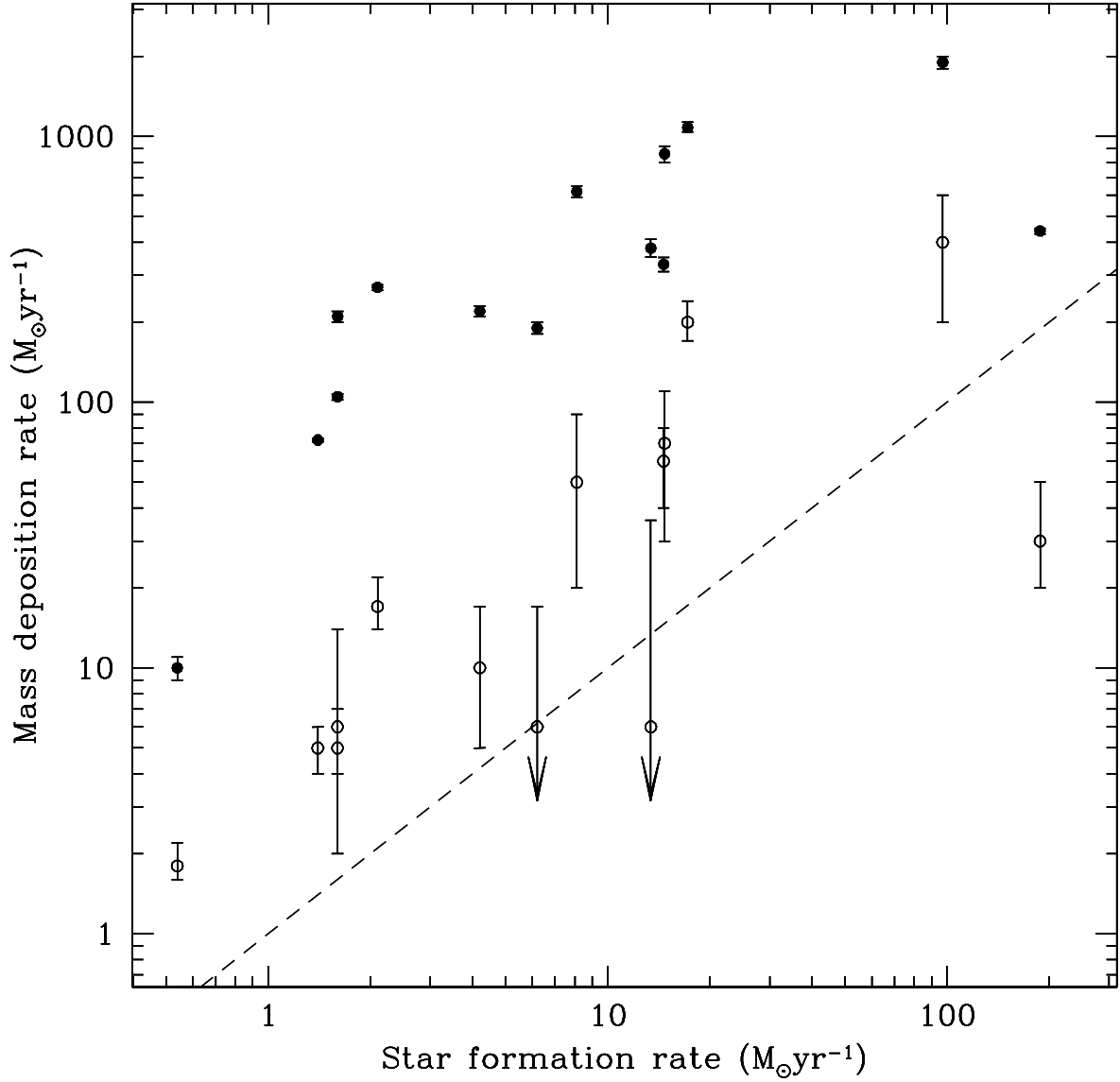}
\caption{
X-ray derived mass deposition rate upper limits against estimated star formation rates.
The closed circles correspond to maximum mass deposition rates, ${\dot M}_I$ if
heating is absent,
and the open circles refer to ${\dot M}_S$, the mass deposition rate
consistent with the X-ray spectra. The dashed line is for equal star
	formation and mass deposition rates.
\label{fig:mdot}}
\end{figure*}

$\dot{M}_S$ was calculated by repeating the spectral
fitting to the annuli within the cooling radius but now adding a cooling flow
model (\texttt{mkcflow}) to the absorbed single-temperature model
(\texttt{phabs(mekal+mkcflow)}).  The XSPEC model \texttt{mkcflow}
models gas cooling between two temperatures and gives the
normalization as a mass deposition rate, $\dot{M}_S$.  For each spectrum, the temperature
of the \texttt{mekal} component was tied to the high temperature of
the \texttt{mkcflow} component and the abundances of the two
components were tied together.	The low temperature of the cooling
flow model was fixed to 0.1 keV.   $\dot{M}_S$ is a measure of the maximum
rate (upper limit) that  gas can be cooling below X-ray temperatures and be consistent 
with  the X-ray spectra.  Detailed spectra
of nearby bright clusters \citep{peterson03,fabian06} tend to show 
an absence of the X-ray coolest gas and indeed
for the inferred $\dot{M}$ to be a function of temperature within a
cluster.  Better quality data for the objects here may lead in some
cases to lower estimates of $\dot{M}_S$.  We have listed the mass deposition
rates for each cluster in Table~\ref{tab:mdot} and plotted these values against star formation 
rates estimated from the infrared luminosity in Figure~\ref{fig:mdot}.


\subsection{Implications}

We see that the SFR is proportional to (but significantly less than) the two estimates 
of mass accretion rate.
The results show that the star formation rate is about 30--100 times
smaller than $\dot{M}_I$, and 3--10 times smaller than $\dot{M}_S$.
The observed trends  between cooling time and the IR luminosity and between
$\dot{M}_S$ and the infrared star formation rates are consistent
with the hypothesis that the cooling ICM is the source of the gas which is forming stars.  
Using a nearly identical approach
to the X-ray  analysis,  \citet{rafferty06} found a similar trend using optical-UV 
star formation rates.   Star formation in these systems
is generally not heavily obscured, and the star formation rates are 
approaching and in some cases agree with the
the cooling upper limits, $\dot{M}_S$.  This is consistent with the
results from X-ray spectroscopy (e.g., \citealt{kaastra01, tamura01, peterson01, peterson03,
peterson06}) which suggests that
most of the gas with short cooling time at a few keV does not cool further. 
Sensitive, high resolution X-ray spectroscopy should soon detect the cooling at the level of
star formation in the Fe XVII lines if this
picture is correct (e.g., \citealt{sanders07b}).  
Our fraction of gas which does cool is a mean number and could be affected
by our  H$\alpha$ selection but we believe the use of a complete X-ray
sample will allow this effect to be quantified.
Nevertheless this number could provide a constraint on the efficiency of
feedback models that prevent the bulk of the ICM from cooling.
If star formation is the ultimate sink for the cooling gas, then the fraction of the 
few keV gas which does cool all the way down should be comparable to the ratio 
SFR/${\dot M}_X$ - which we find to be roughly a few percent. This fraction is comparable
for all the clusters. This suggests that the re-heating mechanism (whatever it is) is very effective 
over a range of size scales and operates nearly all the time (i.e., with a short duty cycle) 
(see \citealt{mcnamara07,peterson06} for reviews).

\subsection{Alternative Energy Sources}

We have proceeded with the assumption that the infrared emission is
solely due to star formation. Here we examine whether there are
reasonable alternative sources of energy for this emission. First we
consider the hot gas, since dust mixing with the gas can be heated and
become a source of mid infrared radiation \citep{dwek86, dwek90}. 
We then consider cosmic ray and other heat sources. 

The hot gas is potentially a rich energy source which could heat the
dust.  A consequence of such heating is the the energy loss from the
gas which means the gas will cool, perhaps even exacerbating the
cooling flow problem. It offers a solution to the problem seen in the
X-ray spectra of cool core clusters in which gas is observed to cool
down to only about 1 keV but no lower \citep{peterson01, tamura01, peterson03}.
The temperature profiles in clusters
mean that the coolest gas is at the smallest radii so if there is {\em
non-radiative} cooling of gas at those radii, say due to mixing with
cold dusty gas, X-ray spectra of the whole core would imply a cooling
flow going down to just 1 keV and appearing to stop, more or less as
observed. This can be seen as 'the missing soft X-ray luminosity'
problem \citep{fabian02}. What is meant by missing soft X-ray
luminosity is the emission missing from a complete cooling flow when
it appears from X-ray spectra to stop at say 1 keV.

\begin{figure*} 
\plotone{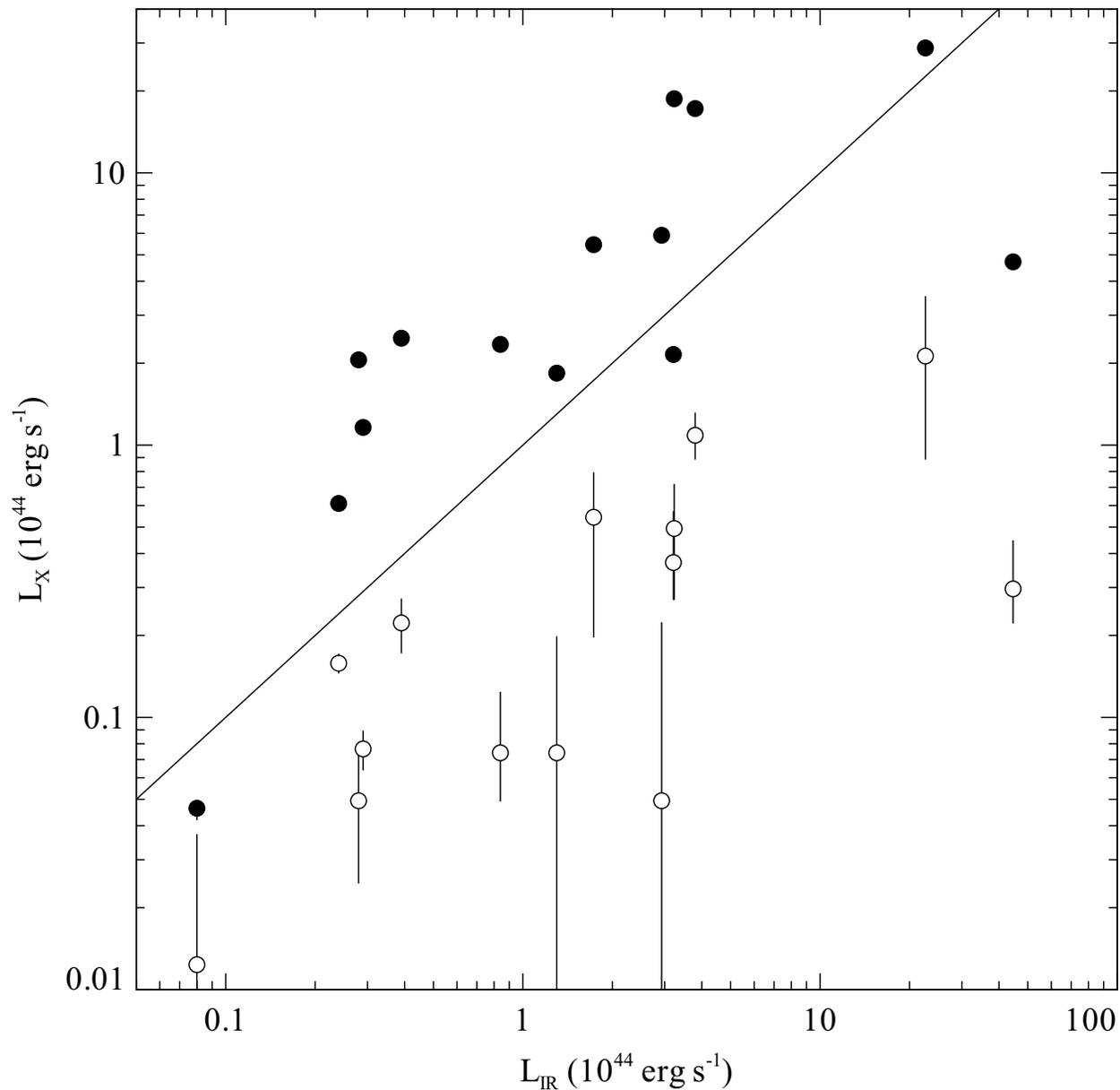}
\caption{
The X-ray luminosity emitted with the cooling radius (a fiducial
radius where the radiative cooling time is 7.7~Gyr, corresponding to
redshift one) is plotted in solid circles against the Spitzer infrared
luminosity. The expected (missing) luminosity emitted below 1~keV by a
continuous cooling flow operating from the cluster virial temperature
to zero K is shown by the open circles. If mixing with dusty cold gas
causes the rapid non-radiative cooling of the intracluster gas below
1~keV then this luminsoity could emerge in the mid-infrared.
\label{fig:alt}}
\end{figure*}

Figure \ref{fig:alt} shows the missing soft X-ray luminosity for our objects
plotted against the IR luminosity as open circles.
This was obtained by fitting the spectra with a cooling flow model which has a 
lower temperature limit of 1~keV.  The missing soft X-ray luminosity is then the 
rate of energy release as that gas cools further to zero K in some non-radiative manner.
There is a correlation, but the normalization misses by about a factor of
5. This means that on average there's 5 times more $L_{\rm IR}$ than needed to
account for non-radiative cooling of the gas below 1~keV.

Better agreement can be obtained by increasing the lower fitted
temperature above 1~keV, but in that case the mass cooling rates rise
from the more modest rates comparable to $\dot M_{\rm S}$ in Fig. ~\ref{fig:mdot} to
the higher, pre-XMM/Chandra values of $\dot M_{\rm I}$.  This is just
because $L_{\rm IR}$ is similar to $L_{\rm cool}$, the luminosity of
the cooling region in the core (i.e.,  where the radiative cooling time is
less than say 5 Gyr), which is shown by the filled circles in Fig.~\ref{fig:alt}.

The result is that if dust mixing in hot gas is the only source of infrared 
emission then we have to face cooling rates much higher than can be 
accommodated in terms of the observed molecular gas \citep{edge01} 
or observed star formation rates (Table ~\ref{tab:SFR_comp}). 
More plausibly, hot gas mixing with 
dusty cold gas is the  source of 10--20 per cent of the infrared emission. 
In this case our results allow for modest rates mass cooling rates of up 
to tens to hundreds $M_\odot$~yr$^{-1}$ comparable to the range shown in 
Table ~\ref{tab:SFR_comp}.

Cosmic rays also fail as an energy source, unless they are recycled.
Since $L_{\rm IR}\sim L_{\rm cool}$ (to within about a factor of 3,
see Fig.\ref{fig:alt}), then the energy required for the infrared is comparable
with the thermal energy within $r_{\rm cool}$. Consequently the cosmic
ray pressure would need to be high with a pressure $P_{\rm CR}=f_{\rm
CR} P_{\rm Th}$ with $f_{\rm CR}>0.3$ and thermal pressure $P_{\rm
Th}$. This is contrary to the quasi-hydrostatic appearance of the
intracluster medium in cluster cores.

Only if there is some efficient mechanism for energy to flow from the
central accretion flow / AGN to the dust can an alternative be viable.
In the absence of any such mechanism, we conclude that the UV
radiation from massive star formation must be the energy source for
the mid-infrared emission measured by Spitzer.


\section{Summary}

\citet{quillen07}  obtained Spitzer photometry of a sample of  62 BCGs in
X-ray bright clusters selected on the basis of BCG H$\alpha$ flux which tends
to favor cool core clusters. They showed that at least half of the BCGs
exhibit an infrared excess with a luminosity $L_{\rm IR} \sim 10^{43} - $ 
few $\times 10^{45}$ ergs/s.  In this paper we examined correlations in the 
data  and discussed implications for cool core clusters. 

BCGs with an IR excess are found mainly in clusters at high X-ray luminosity 
($L_{\rm X} > 10^{44}$
ergs s\mone). But not all high $L_{\rm X}$ clusters have 
a BCG with an IR excess. 

The IR luminosity is proportional to the H$\alpha$ luminosity, suggesting that they
are powered by the same or a related source of energy. We suggest that star formation
is the dominant power source for the the IR and H$\alpha$ emission.
The H$\alpha$ luminosity falls below the \citet{kennicutt98} relation probably because
the spectroscopic apertures exclude much of the extended emission line nebulae. 
The inferred star formation rates estimated from the IR luminosity are in the range 
about 1 to 50 
$M_\odot$~yr$^{-1}$. 
In most BCGs, supernovae produced by star 
formation with this SFR cannot account for the X-ray luminosity and so cannot 
be responsible for re-heating the ICM. 

The radio emission in the BCG is dominated by that produced by an AGN rather than 
star formation.  However, there is a modest correlation between radio and IR 
emission. This suggests the feeding of the AGN and the fueling of the star 
formation may have a common origin, perhaps gas cooling from the hot ICM. 

The mass of molecular gas (estimated from CO observations) is correlated with the IR 
luminosity as found for normal star forming galaxies. The gas depletion time scale is about
1 Gyr. Given that clusters are relatively young (perhaps 4-6 Gyr since
last major merger) then it is possible that there may have been insufficient time for a
complete  steady-state (cooling leads to cold gas leads to star formation) to be set up.

We fit a Schmitt-Kennicutt relation to the molecular gas mass vs. SFR and estimate a 
rough star forming region diameter. For most BCGs the implied sizes of 10-20 kpc is 
comparable to that of the color variations seen in the IRAC data and to
the optical emission line nebulae. This is  consistent with the hypothesis that the star 
formation occurs in molecular gas associated with the emission line nebulae and with
evidence that the emission line nebulae are mainly powered by ongoing star formation. 

BCGs in clusters with shorter cooling times for the hot ICM have higher IR luminosities. 
We find a strong correlation between mass deposition rates (${\dot M}_X$) estimated from
the X-ray emission and the SFR.  
The star formation rate is about 30--100 times
smaller than $\dot{M}_I$ - the mass accretion rate derived from imaging, and 3--10 
times smaller than $\dot{M}_S$ - the rate derived from spectroscopy.
The observed trends  between cooling time and the IR luminosity and between
$\dot{M}_S$ and the infrared star formation rates are consistent
with the hypothesis that the cooling ICM is the source of the gas which is forming stars.  
The correlation between mass deposition rates estimated
from the X-ray radiative losses and the star formation rates
suggest that the fraction of gas that does cool is set
by the balance of heating and cooling by the cooling flow. The low value of
SFR/(${\dot M}_X$) suggests that heating is likely to be very efficient in
preventing most of the gas a temperatures of a few keV from
cooling further.

\acknowledgments
This work is based in part on observations made with the Spitzer Space 
Telescope, which is operated by the Jet Propulsion Laboratory, California 
Institute of Technology under a contract with NASA. Support for this work 
at University of Rochester and Rochester Institute of Technology was provided 
by NASA through an award issued by JPL/Caltech. We are grateful to the 
referee for helpful comments. 

{}
\clearpage

\begin{deluxetable}{lcrr}
\tablewidth{0pt}
\tablecaption{Spearman Rank-Order Correlation Coefficients\label{tab:corr}}
\tablehead{
\colhead{Plot name} &
\colhead{Figure number} &
\colhead{Correlation Coefficient} &
\colhead{Two-sided significance }
}
\startdata
$L_X$ vs $L_{\rm IR}$  & \ref{fig:lxvslir}	&  0.63 & $5.0\times10^{-5}$ \\
$F_X$ vs $F_{\rm IR}$  &                  	&  0.14 & 0.40 \\
$L_X$ vs 8/5 & \ref{fig:lxvs85} &   0.38 & $3\times10^{-3}$ \\
$L_{1.4 \rm GHz}$ vs $L_{\rm IR}$ & \ref{fig:lradiolir} & 0.41	& 0.02    \\
F$_{1.4 \rm GHz}$ vs $F_{\rm IR}$ &                    & -0.09	& 0.61    \\
$L_{\rm H\alpha}$ vs $L_{\rm IR}$ & \ref{fig:lalphalir} & 0.91 & $3.6\times10^{-12}$ \\
$F_{\rm H\alpha}$ vs $F_{\rm IR}$ &                     & 0.65 & $1.1\times10^{-4}$ \\
$L({\rm H}\alpha)$ vs $L(24\mu {\rm m})$ &                  & 0.84 & $2 \times 10^{-15}$ \\
$F({\rm H}\alpha)$ vs $F(24\mu {\rm m})$ & \ref{fig:havs24} & 0.67 & $4 \times 10^{-8}$ \\
M(H$_2$) vs $L_{\rm IR}$ & \ref{fig:coplots1} &  0.95  & $1.3\times10^{-10}$ \\
F(CO)    vs $F_{\rm IR}$ &                    &  0.81  & $1.7\times10^{-5}$ \\
%
\enddata
\tablecomments{Col 1. Correlation being tested. 
Col 2. Figure which plots the data. 
Col 3. Spearman Rank-Order Correlation Coefficients. 
Col 4. Two-sided significance of the 
correlation's deviation from zero. 
The most signficant correlations are that between H$_\alpha$ and infrared luminosity
and that between molecular gas mass and infrared luminosity.
Most correlations are done both on fluxes and luminosities.
}
\end{deluxetable}


\begin{deluxetable}{lcc}
\tablewidth{0pt}
\tablecaption{Star Formation Rate(SFR) 
\label{tab:selected}}
\tablehead{
\colhead{Cluster} &
\colhead{L$_{IR}$} &
\colhead{SFR}
\\
\colhead{}&
\colhead{($10^{44}$erg s$^{-1}$)} &
\colhead{($M_{\odot}$/yr)}
}
\startdata
Z2089*    & 64.68 & 271  \\
A2146*    & 45.46 & 192  \\
A1068*    & 44.61 & 188  \\ 
R0821+07* & 8.47  & 37   \\ 
\hline
R1532+30* & 22.62 & 97   \\
Z8193*    & 13.70 & 59   \\
Z0348*    & 11.92 & 52   \\
A0011*    & 7.97  & 35   \\
PKS0745-1 & 3.80  & 17.2 \\
A1664     & 3.21 & 14.6  \\
R0352+19  & 2.40 & 11.1  \\
NGC4104   & 0.80 & 4.0   \\
R0338+09  & 0.39 & 2.1   \\ 
\hline
R0439+05* & 4.17 & 18.7  \\
A2204    & 3.23 & 14.7   \\
A2627    & 1.59 & 7.5    \\
A0115    & 1.30 & 6.2    \\
Z8197    & 0.72 & 3.6    \\ 
\hline
R2129+00 & 2.93 & 13.4      \\
A1204    & 1.73 & 8.1       \\
A0646    & 1.49 & 7.1       \\
A2055    & 1.46 & 7.0       \\
A0291    & 1.30 & 6.3       \\
A1885    & 1.04 & 5.1       \\
A3112    & 0.84 & 4.2       \\
A2292    & 0.80 & 4.0       \\
A1930    & 0.75 & 3.8       \\
Z8276    & 0.74 & 3.7       \\
A4095    & 0.29 & 1.6       \\
A0085    & 0.28 & 1.6       \\
A2052    & 0.24 & 1.4       \\
R0000+08 & 0.20 & 1.2       \\
NGC6338  & 0.18 & 1.0       \\
R0751+50 & 0.10 & 0.65      \\
A0262   &  0.08 & 0.54    
\enddata
\tablecomments{
Infrared luminosities are from \citet{quillen07} 
estimated from the 15$\mu$m wavelength
for BCGs that are detected at 70$\mu$m or have color ratio
$F_{8 \mu {\rm m}}/F_{5.8 \mu {\rm m}}>1.0$ or 
$F_{24 \mu {\rm m}}/F_{8 \mu {\rm m}}>1.0$.
The star formation rate was estimated using Equation \ref{eqn:SFR}.
The top section contains four BCGs that are suspected
to harbor dusty Type II AGNs. Z2089, A2146 and A1068 exhibit a
red $F_{4.5 \mu {\rm m}}/F_{3.6\mu {\rm m}} $ color and all four
exhibit high $[$OIII$]$(5007)/H$\beta$ flux ratios.
Note that if there is an AGN present in one of these clusters,
the SFR may be overestimated. 
The second set is
the remaining 10 BCGs with $F_{8 \mu {\rm m}}/F_{5.8 \mu {\rm m}}>1.3$.
The third section is the set of 6 clusters with
$1.0<F_{8 \mu {\rm m}}/F_{5.8 \mu {\rm m}}<1.3$.
The fourth set is the remaining BCGs with IR excesses.
Specifically they have ratios
$F_{8 \mu {\rm m}}/F_{5.8 \mu {\rm m}}>1.0$,
$F_{24 \mu {\rm m}}/F_{8.0 \mu {\rm m}}>1.0$,
or a detected 70$\mu$m flux.
The BCGs marked with a $*$ can be classified as LIRGs
since they have L$_{IR}$ greater than $10^{11}L_\odot$.
Objects with $F_{8 \mu {\rm m}}/F_{5.8 \mu {\rm m}}<1.0$ or 
$F_{24\mu{\rm m}}/F_{8\mu {\rm m}} < 1$ and lacking a $70\mu$m detection 
are listed in Table 3 of paper I \citep{quillen07}  with upper limits
on $L_{IR}$.  For these
objects $L_{IR} \la 0.3 \times 10^{44}$erg~s$^{-1}$ and 
corresponding star formation
rates are lower than $\la 1 M_\odot$~yr$^{-1}$.
}
\end{deluxetable}

\begin{deluxetable}{lcrrrrrrrrrr}
\tabletypesize{\scriptsize}
\setlength{\tabcolsep}{0.02in}
\rotate
\tablecaption{Comparison of Estimates of Star Formation Rate \label{tab:SFR_comp}}
\tablehead{
\colhead{BCG } &
\colhead{OD07} &
\colhead{C99} &
\colhead{HM05 }& 
\colhead{MO93} &
\colhead{M95} &
\colhead{M04 }& 
\colhead{M05} &
\colhead{B03} &
\colhead{MO89 }& 
\colhead{OD04 }&
\colhead{D07} 
}
\startdata
%
A262   & 0.5 &    &      & 0.02 &     &     &     &     &      &      &     \\ 
A2597  &     &    &      &  12  &     &     &     &     &      &  10  &  4   \\
A1795  &     &  2 &   9  &  12  &     &     &     &     &  1.8 &  10  &      \\
A1835  &     & 77-125& 123&     &     &     & 100 &     &      &      &      \\
 "     &     &    &      &      &     &     & 138 (FIR) & &     &      &      \\
HydA   &     &    &  9.5 &      & 1 (b)   &     &     &     &      &      &      \\
 "     &     &    &      &      & 23-35 (c) &   &     &     &      &      &      \\  
A2052  & 1.4 & 0.96 &    &      &     &     &     &0.4-0.6& 0.16&     &      \\   
A1068  & 188 &  30 &     &      &     &16-40&    &     &      &      &      \\
 "     &     &    &      &      &     & 68 (IR)&   &     &     &       &     \\ 
A1664  &  14 &  23 &     &      &     &      &    &     &     &       &     \\
R1532  &  97 &  12 &     &      &     &     &     &     &     &       &     \\
A2146  & 192 &  5.6 &    &      &     &     &     &     &     &       &     \\          
PKS0745 & 17 &    & 129  &      &     &     &     &     &     &       &     \\     
\enddata
\tablecomments{Comparison of star formation rates from this paper (Col 2) with estimates
from the literature (Col 3-12) Values are in units of $M_\odot$ yr$^{-1}$. References: 
OD07 =  this paper, C99  =  Crawford et al. (1999), HM05    =  Hicks \& Mushotzky (2005),
       MO93    =  McNamara \& O'Connell (1993), M95     =  McNamara (1995), 
       M04     =  McNamara et al. (2004), M05     =  McNamara et al. (2005),
       B03     =  Blanton et al. (2003), MO89    =  McNamara \& O'Connell (1989),
       OD04 =  O'Dea et al. (2004), D07   =  Donahue et al. (2007).
For Hydra A, (b) and (c) refer to short duration burst and continuous 
star formation models, respectively. 
}
\end{deluxetable}

\begin{deluxetable}{lcrr}
\tablewidth{0pt}
\tablecaption{Mass Deposition Rates\label{tab:mdot}}
\tablehead{
\colhead{Cluster} &
\colhead{$\dot{M}_S$} & 
\colhead{$\dot{M}_I$} \\
\colhead{} &
\colhead{($M_\odot$/yr)} &
\colhead{($M_\odot$/yr)}
}
\startdata
A1068     & 30$^{+20}_{-10}$    & 440$^{+10}_{-10}$ \\
R1532+30  & 400$^{+200}_{-200}$ & 1900$^{+100}_{-100}$ \\
PKS0745-1 & 200$^{+40}_{-30}$   & 1080$^{+50}_{-40}$ \\
A1664     & 60$^{+20}_{-20}$    & 330$^{+20}_{-20}$ \\
R0338+09  & 17$^{+5}_{-3}$      & 270$^{+6}_{-6}$ \\
A2204     & 70$^{+40}_{-40}$    & 860$^{+60}_{-60}$ \\
A115      & 6$^{+11}_{-6}$      & 190$^{+10}_{-10}$ \\
R2129+00  & 6$^{+30}_{-6}$      & 380$^{+30}_{-20}$ \\
A1204     & 50$^{+40}_{-30}$    & 620$^{+30}_{-30}$ \\
A3112     & 10$^{+7}_{-5}$      & 220$^{+10}_{-10}$ \\
A4059     & 5$^{+2}_{-1}$       & 105$^{+2}_{-3}$ \\
A0085     & 6$^{+8}_{-4}$       & 210$^{+10}_{-10}$ \\
A2052     & 5$^{+1}_{-1}$       & 72$^{+1}_{-1}$ \\
A0262     & 1.8$^{+0.4}_{-0.2}$  & 10$^{+1}_{-1}$ \\
\enddata
\tablecomments{Mass deposition rates calculated within $r_{cool}$ using Chandra
  and XMM data.  $\dot{M}_S$ is a measure of the mass deposition rate
  consistent with the X-ray spectra and $\dot{M}_I$ is a measure of the mass deposition rate
  if heating is absent. Sources are listed in order of decreasing SFR. 
}
\end{deluxetable}

\end{document}